\documentclass[conference]{IEEEtran}

\usepackage{epsfig}
\usepackage{graphicx}
\usepackage{amsmath}
\usepackage{amssymb}
\usepackage{amsxtra}
\usepackage{here}
\usepackage{rawfonts}
\usepackage{times}
\usepackage{url}
\usepackage{cite}
\usepackage{cases}
\usepackage{adjustbox}

\usepackage{slashbox}
\usepackage{epstopdf}

\usepackage{cite}

\newlength{\aligntop}
\newlength{\alignbot}
\makeatletter

\def\ps@IEEEtitlepagestyle{%
\def\@oddfoot{\mycopyrightnotice}%
\def\@evenfoot{}%
}
\def\mycopyrightnotice{%
{\footnotesize 978-1-5090-1164-3/16/\$31.00 \textcopyright2016 IEEE\hfill}% <--- Change here
\gdef\mycopyrightnotice{}% just in case
}
\renewenvironment{align}{%
  \vspace{\aligntop}
  \start@align\@ne\st@rredfalse\m@ne
}{%
  \math@cr \black@\totwidth@
  \egroup
  \ifingather@
    \restorealignstate@
    \egroup
    \nonumber
    \ifnum0=`{\fi\iffalse}\fi
  \else
    $$%
  \fi
  \ignorespacesafterend%
  \vspace{\alignbot}\par\noindent
}
\IEEEoverridecommandlockouts

\begin{document}
\title{\LARGE On Bounded Rationality in Cyber-Physical Systems Security: Game-Theoretic Analysis with Application to Smart Grid Protection\vspace{-0.55cm}}
\author{\IEEEauthorblockN{Anibal Sanjab$^1$ and Walid Saad$^1$} \IEEEauthorblockA{\small
$^1$ Wireless@VT, Bradley Department of Electrical and Computer Engineering, Virginia Tech, Blacksburg, VA, USA,\\
 Emails: \url{{anibals,walids}@vt.edu}\vspace{-0.6cm}
 }%
%\thanks{This research was supported by the U.S. National Science Foundation under Grant CNS-1446621.}
\thanks{This work was supported by the U.S. National Science Foundation under Grants ECCS-1549894 and CNS-1446621.}
    }
\date{}
\maketitle

\begin{abstract}
%Cyber-physical systems (CPS) represent a combined system with tight interconnection between the physical layer and its underlying information and communication layers. Even though this interconnection promotes traditional physical system into more intelligent and efficient ones, it introduces various security risks. 
In this paper, a general model for cyber-physical systems (CPSs), that captures the diffusion of attacks from the cyber layer to the physical system, is studied. In particular, a game-theoretic approach is proposed to analyze the interactions between one defender and one attacker over a CPS. In this game, the attacker launches cyber attacks on a number of cyber components of the CPS to maximize the potential harm to the physical system while the system operator chooses to defend a number of cyber nodes to thwart the attacks and minimize potential damage to the physical side. The proposed game explicitly accounts for the fact that both attacker and defender can have different computational capabilities and disparate levels of knowledge of the system.
%Given that the attacker and defender can have different level of knowledge of the system and different computational abilities, they might act sub-optimally, i.e. with bounded rationality. 
To capture such \emph{bounded rationality} of attacker and defender, a novel approach inspired from the behavioral framework of cognitive hierarchy theory is developed. In this framework, the defender is assumed to be faced with an attacker that can have different possible thinking levels reflecting its knowledge of the system and computational capabilities. To solve the game, the optimal strategies of each attacker type are characterized and the optimal response of the defender facing these different types is computed. This general approach is applied to smart grid security considering wide area protection with energy markets implications. %Numerical results have been derived through an aplication to the PJM 5-bus test system. 
Numerical results show that a deviation from the Nash equilibrium strategy is beneficial when the bounded rationality of the attacker is considered. Moreover, the results show that the defender's incentive to deviate from the Nash equilibrium decreases when faced with an attacker that has high computational ability.  
\end{abstract}
%\vspace{-0.25cm}
\section{Introduction}
Cyber-physical systems (CPSs) are characterized by a tight interconnection between the physical system and its underlying information and communication layers. Such a cyber-physical interconnection introduces new security risks that significantly differ from classical networked systems~\cite{DefCPS1,DefCPS2,Poor}. In fact, many reports have discussed the vulnerability of a CPS to attacks and the destructive effect that such attacks can have~\cite{WallStreet,CNN}.

CPS security has been the subject of various research works recently~\cite{SecCtrlBerkeley,AttDetDorfler,SecCPSBasar,ResilientCtrlBasar,CascadingFailure}. The work in~\cite{SecCtrlBerkeley} characterizes a number of sought CPS operational objectives under various potential security threats. In~\cite{AttDetDorfler}, the authors propose a mathematical framework for modeling and detection of various types of attacks on a CPS. The authors in~\cite{SecCPSBasar} propose a hierarchical security architecture of CPS and develop a cross layer approach to devise security solutions against attacks. Resilient control against denial-of-service is studied in~\cite{ResilientCtrlBasar}. The work in~\cite{CascadingFailure} considers cascading failures due to malicious attacks and devises a robust defense strategy to protect the CPS against random attacks.  

These research efforts have focused on general CPS security problems that can be applicable to various CPS fields such as the power grid and transportation systems. %(e.g. smart grids, transportation systems, water distribution, etc.). 
Meanwhile, several recent works have focused on studying security threats tailored to a specific domain, with a particular focus on the smart grid. Securing the smart grid has indeed attracted significant attentions such as in~\cite{Poor, SecTechSG} and~\cite{CPSSmrtGridProceedings}. In~\cite{Poor}, the authors consider data injection attacks on smart grids and devise a defense policy to outcast such attacks. The work in~\cite{SecTechSG} provides an analysis of various security technologies to improve the robustness of the smart grid against attacks. Along the same lines, the authors in~\cite{CPSSmrtGridProceedings} analyze various cyber attacks on the smart grid and propose solutions for improving its security.   

While all these works provide interesting contributions to the field of CPS and smart grid security, only few of them consider the underlying strategic interaction between attackers and defenders. %when devising optimal strategies for defense and attack. 
In fact, an attacker typically aims at choosing an attack strategy that would cause a damage to the system while the defender aims at decreasing that damage caused by the attack. Thus, the actions and objectives of a CPS attacker and defender are intertwined which makes the study of their strategic interaction necessary. Moreover, in the works that consider this strategic behavior, as in~\cite{SecCPSBasar} and~\cite{ResilientCtrlBasar}, it is assumed that attackers and defenders are fully rational players that can act optimally and strategically. However, when faced with risks, incomplete information, stress, and constraints such as limited time to act and high involved complexity, as is the case of CPS security, individuals tend to act with limited rationality~\cite{ThinkingFastandSlowKahneman}. Thus, such \emph{bounded rationality} needs to be explicitly accounted for when analyzing the strategic interaction of the players within CPS security situations for a better modeling of the attackers' and defenders' actual behaviors.  %In fact, many CPS security decisions are driven by individuals, such as hackers and administrators, which take their decisions under paramount amount of stress and complexity and hence may act with bounded rationality. %To our knowledge very few papers have considered deviation from full rationality~\cite{PoorSecGames}. For example, in~\cite{PoorSecGames}, one of the players in a security game is assumed to act as a follower, in some instances, and as a non-follower in others which showcases possible deviations from the well known Stackelberg equilibrium that is applied in a number of security situations. 

The main contribution of this paper is to develop a novel framework for analyzing CPS and smart grid security, in the presence of an attacker and defender with bounded rationality. Given a general CPS model, the problem is formulated as a zero-sum game between the attacker and defender that are interacting over the cyber side of the CPS. In this game, the attacker aims at launching cyber attacks on a number of cyber nodes of the CPS to damage some of the physical components by capitalizing on the diffusion of failures from the cyber to the physical components. In contrast, the defender aims at defending a number of cyber nodes to stop such attacks. Since the attacker and defender can have different computational abilities and levels of knowledge of the CPS, we propose a novel approach to capture such \emph{bounded rationality} inspired by the behavioral framework of cognitive hierarchy theory~\cite{CH01}. Our proposed framework considers that the defender can be faced, in practical situations, with an attacker possessing one of various possible levels of computational abilities and knowledge depth. Thus, choosing a defense strategy while always assuming that the attacker is a very complex strategic thinker, as in conventional games~\cite{SecCPSBasar,ResilientCtrlBasar}, is not always optimal. Hence, our bounded rationality framework assumes that the defender can be faced with an attacker than can have one of many \emph{levels of thinking} reflecting the complexity of its used strategy. 
To solve this game, we characterize the various levels of thinking of the attacker and accordingly derive its optimal strategy. %based on its level, and derive the optimal defense strategy of the defender considering probable attacker's type. %when faced wit. Moreover, the defender is considered to be faced with an attacker which can have one of many levels of thinkings with a certain probability distribution. 
Moreover, the optimal strategy of the defender is chosen to be the one that maximizes its expected payoff given a probability distribution of the attacker's thinking levels. 

As a case study, we consider a wide area protection scenario in the smart grid. In this study, we focus on the economic effects that a false disconnection of a transmission line can have on the system. Our numerical results over the PJM-5 bus system show that the defender can have an incentive to deviate from its Nash equilibrium strategy knowing that the attacker can be acting with bounded rationality. Moreover, these results also showcase the effect that the probability of facing each attacker level has on the optimal attack and defense strategies.     

The rest of this paper is organized as follows. Section~\ref{sec:CPSSecModel} introduces our general CPS security model as well as the proposed game with bounded rationality.  Section~\ref{sec:WideAreaProtectionandMarkets} presents a case study focusing on wide area protection of the smart grid illustrating our proposed model and game while Section~\ref{sec:Conclusion} draws some conclusions.  %\vspace{-0.4cm}
\section{System Model and Game Formulation}\label{sec:CPSSecModel}
%\subsection{CPS Security Model}\label{subsec:CPSInterconnection}
\begin{figure}[t!]
  \begin{center}
   \vspace{-0.35cm}
    \includegraphics[width=6.4cm]{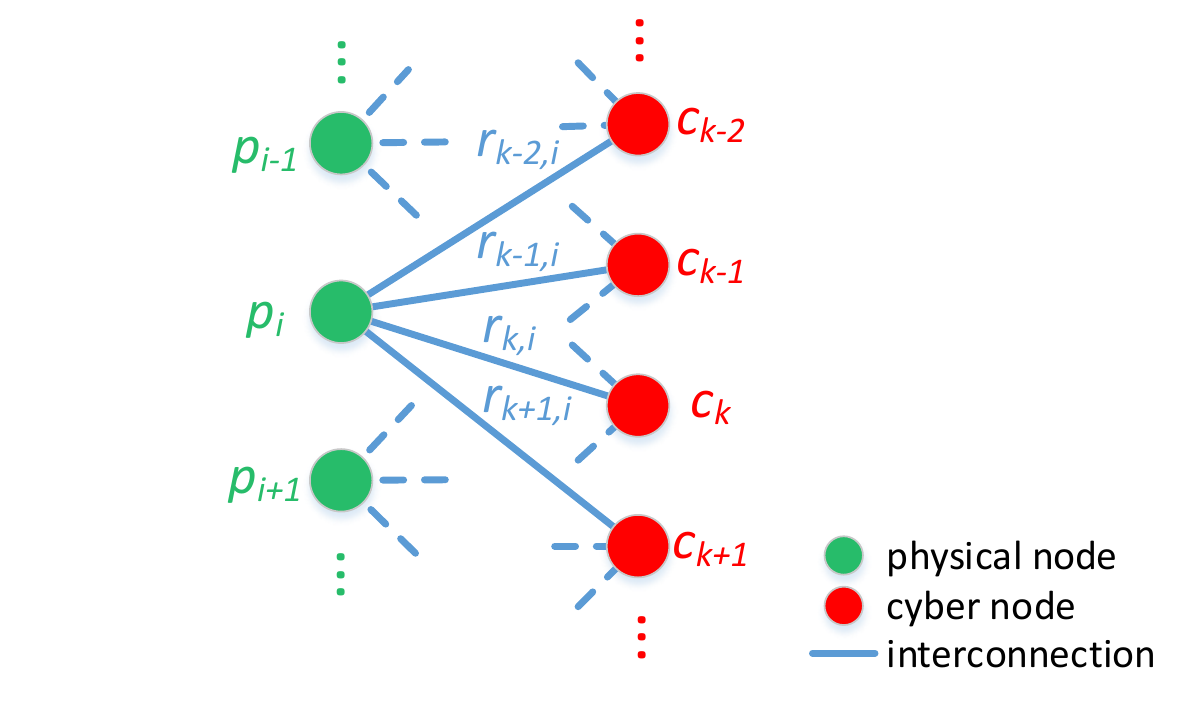}
    \vspace{-0.5cm}
    \caption{\label{fig:CPSModel} Cyber-Physical Interconnection.}
  \end{center}\vspace{-0.9cm}
\end{figure}
Consider a CPS composed of $N_c$ cyber and $N_p$ physical nodes that are strongly interdependent. Let $\mathcal{C}$ and $\mathcal{P}$ be, respectively, the sets of cyber and physical nodes. In this model, security breaches can spread from the cyber to the physical realms. As illustrated in Fig.~\ref{fig:CPSModel}, we let $r_{c,p}$ be the interconnection between a cyber node $c\in\mathcal{C}$ and a physical node $p\in\mathcal{P}$. In fact, control laws governing the operation of the physical system depend on local and remote data collected by cyber nodes. Such data is sent via communication channels to a supervisory control and data acquisition system (SCADA) which, in turn, sends control signals back to the cyber nodes initiating a control action over the physical nodes. In this regard, $r_{c,p}$ is a weight that captures the effect of the data sent by cyber node $c$ on the control action over physical node $p$. Accordingly, from a security perspective, $r_{c,p}$ represents the probability that $p$ fails due to corrupt data sent by $c$. Hereinafter, we use ``failed node'' to refer to a cyber node sending corrupt data. This failure of $c$ can be due to a cyber attack on this node or to other reasons such as a software bug or a misconfiguration. Hence, $r_{c,p}$ can be expressed as $r_{c,p}=\textrm{Pr}(p \,\textrm{fails}\,|\,c \,\textrm{has failed})$, while $\sum_{c\in\mathcal{C}}r_{c,p}=1$.  
%\begin{align}\label{eq:rik}
%r_{k,i}=Pr(p_i \,\textrm{fails}/c_k \,\textrm{has failed}),
%\end{align} 
 
Let $\pi_p$ be the probability of failure of $p\in\mathcal{P}$ %, for $i=1,...,N_p$, 
due to failures of a number of cyber nodes, and $\kappa_c$ the probability of failure of one of the cyber nodes $c\in\mathcal{C}$. Accordingly, %for $k=1,...,N_c$, 
%The following expression can hence be derived: 
$\pi_p$ will be given by %$\pi_p=\sum_{c=1}^{N_c}r_{c,p}\kappa_c.$ 
\begin{align}\label{eq:PrFailure}
\pi_p=\sum_{c=1}^{N_c}r_{c,p}\kappa_c.
%\pi_i^p=\sum_{k=1}^{N_c}r_{k,i}\pi_k^c.
\end{align}\vspace{+0.01cm}

We denote by $\boldsymbol{R}$ the matrix of interconnections between cyber and physical nodes, and $\boldsymbol{\pi}=[\pi_1,...,\pi_{N_p}] \in\mathbb{R}^{N_p}$ and $\boldsymbol{\kappa}=[\kappa_1,...,\kappa_{N_c}]\in\mathbb{R}^{N_c}$ the failure probability vectors of the physical and cyber nodes, respectively. Accordingly,~(\ref{eq:PrFailure}) can be rewritten in matrix form as follows:%\vspace{-0.2cm}
\begin{align}\label{eq:RiskMatrix}
\boldsymbol{\pi}=\boldsymbol{\kappa}\boldsymbol{R}.
\end{align} \vspace{-0.6cm}

Each physical node $p$ is associated with a cost of failure $f_p$. %representing the cost of loosing this physical component. %Let $\boldsymbol{f}^p$ be the vector of of costs associated with the loss of every physical component. 
Hence, the expected total loss to the system, $E_f$, is given by: %\vspace{-0.2cm}
\begin{align}\label{eq:ExpectedLoss}
E_f=\sum_{p=1}^{N_p}\pi_p f_p.
\end{align}
%\vspace{-0.8cm}
\subsection{Game Formulation}\label{subsec:GameFormulation}

In the absence of attacks, the probability of failure of each cyber node is typically small and is only due to the presence of software bugs or misconfiguration by, for example, maintenance personnel. %leading to its failure. 
Thus, under no cyber attacks $\boldsymbol{\kappa}\approx\boldsymbol{0}$ and, thus, $\boldsymbol{\pi}\approx\boldsymbol{0}$. However, when a cyber node $c$ is attacked, the probability of failure of this node goes up to $\kappa_c=1$. As a result, as seen from~(\ref{eq:RiskMatrix}), this attack increases the risk of failure of the physical components that are interconnected to $c$ thus increasing $E_f$ as per~(\ref{eq:ExpectedLoss}). %For example, consider the case in which the pre-attack failure probabilities of the cyber and physical components be, respectively, $\boldsymbol{\pi}^{c,o}=0$ and $\boldsymbol{\pi}^{p,o}$ represent the pre-attack failure probabilities of, respectively, the cyber and physical components. An attack on $c_k$ will lead to:
On the other hand, defending $c$ protects it from failures in which case $\kappa_c=0$ even when $c_k$ is attacked\footnote{Attack and defense are assumed to be always successful such that an attack on $c$ will certainly lead to its failure while defending $c$ leads to its non-failure.}. For example, if an attacker induces a malware over a cyber node, when this node is defended, the malware is detected and eliminated. 

The attacker hence aims at maximizing $E_f$ while the defender, which is the system operator, aims at minimizing it. %(or equivalently maximize $-E_f$). 
%This gives rise to a conflicting strategic interaction between the attacker and defender which we model through the use of game theory~\cite{GT01}. To this end, to 
To analyze the optimal decision making of each of the attacker and defender, we formulate a noncooperative zero-sum game~\cite{GT01} $\Xi=\langle \mathcal{I}, (\mathcal{S}_{i})_{i\in\mathcal{I}}, (U_{i})_{i\in\mathcal{I}}\rangle$. Here, $\mathcal{I}=\{d,a\}$ is the set of players: defender ($d$) and attacker ($a$). $\mathcal{S}_i$ is the set of actions available to player $i \in \mathcal{I}$ which consists of choosing a subset of cyber nodes to defend or attack. Let $n_d$ and $n_a$ be the number of nodes that can be, respectively, defended by the defender and attacked by the attacker. Then, we have $|\mathcal{S}_d|=\big(_{n_d}^{N_c}\big)$ and $|\mathcal{S}_a|=\big(_{n_a}^{N_c}\big)$. %where $\big(_{n_x}^{N_y}\big)$ denotes the number of $n_x$-combinations out of $N_y$. 
$U_i$ is the utility function of player $i\in\mathcal{I}$ and is such that for $s_i\in\mathcal{S}_i$, 
\begin{align}\label{eq:Utility}
U_d(s_d,s_a)=-U_a(s_d,s_a)=-E_f, 
\end{align}%\vspace{-0.1cm}
where $E_f$ is given by~(\ref{eq:ExpectedLoss}). 

\subsection{Solution Concept}\label{subsec:SolutionConcept}

The most commonly adopted equilibrium concept for such static noncooperative games is the Nash equilibrium (NE)~\cite{GT01}. %A NE consists of a strategy profile of the game from which none of the players has any incentive to deviate. 
In this regard, let $\gamma_i\in\Gamma_i$ be a probability distribution over the strategy set of player $i$ where ${\Gamma}_i$ is the set of all possible such distributions. Thus, $\gamma_i(s)$ represents the probability of player $i$ choosing strategy $s\in\mathcal{S}_i$ while $\sum_{s\in\mathcal{S}_i}\gamma_i(s)=1$. Accordingly, each player's expected utility is given by:
%{\small
\begin{flalign}\label{eq:ExpectedU}
\bar{U}_{d}(\boldsymbol{\gamma}_d,\boldsymbol{\gamma}_a)&=-\bar{U}_{a}(\boldsymbol{\gamma}_d,\boldsymbol{\gamma}_a)\nonumber\\
&=-\sum_{s_d\in\mathcal{S}_d}\sum_{s_a\in\mathcal{S}_a}\gamma_d(s_d)\gamma_a(s_a)U_a(s_d,s_a).  
\end{flalign}%}

A \emph{best response strategy} of a rational player $i$, $\gamma^*_i$, is one that maximizes its expected utility facing its opponent's strategy, $\gamma_{-i}$:\vspace{-0.2cm} %That is, 
\begin{align}\label{eq:BestResponse}
\bar{U}_i(\gamma^*_i,\gamma_{-i})\geq\bar{U}_i(\gamma_i,\gamma_{-i}) \>\forall \gamma_i\in\Gamma_i.
\end{align}

When every player plays a best response strategy against its opponent's strategy, the game reaches an equilibrium. %Namely, none of the players have any incentive to deviate from this equilibrium since each is playing the strategy that maximize its utility facing the strategy played by the opponent. 
Thus, the strategy profile $(\gamma^*_i,\gamma^*_{-i})$ is a NE of the game when $\forall i\in\mathcal{I}$~\cite{GT01}:\vspace{-0.2cm}
\begin{align}\label{eq:MSNE}
\bar{U}_i(\gamma^*_i,\gamma^*_{-i})\geq \bar{U}_i(\gamma_i,\gamma^*_{-i}) \>\forall \gamma_i\in\Gamma_i.
\end{align}

%Since in this equilibrium a player randomize its choice over its action space rather than choosing one action with complete certitude, this equilibrium concept is known as a mixed strategy Nash equilibrium (MSNE).
%\begin{remark}
%Since $\gamma^*_i$ consists of an optimal distribution over the action space of $i$, $i$ is known to mix between its strategies and the underlying equilibrium is known as a mixed strategy Nash equilibrium (MSNE).
%%An equilibrium in which each player $i$ chooses a unique optimal strategy $s^{i$ with a probability $1$%, i.e. $\gamma_i(s_i)=1$ and $\gamma_i(s'_i)=0\, \forall s'_i\in\mathcal{S}_i$$\setminus$$s_i$, 
%%$i$ is known to have chosen a pure strategy. Whereas, when $i$ chooses $\gamma_i$ that randomizes between its strategies, $\gamma_i$ is referred to as a mixed strategy. When the equilibrium strategies of all players are pure strategies, this equilibrium is known as a pure strategy Nash equilibrium (PSNE). Otherwise, the equilibrium of the game is referred to as a mixed strategy Nash equilibrium (MSNE).  
%\end{remark}

\subsection{Notion of Bounded Rationality}\label{subsec:BoundedRationality}
 
The NE as defined in~(\ref{eq:MSNE}) assumes that both players are \emph{strategic thinkers, act rationally, and have complete knowledge of the game}. However, when faced with risk and uncertainty,  individuals are known to deviate from full rational behavior\footnote{Even in the case of automated attack and defense, the high required computational ability and short time to act can lead to taking sub-optimal decisions.}~\cite{ThinkingFastandSlowKahneman}.  

In fact,~(\ref{eq:MSNE}) requires every player to anticipate the exact cost, $f_p$, caused to the system due to the loss of each physical component $p\in\mathcal{P}$. With this knowledge, the attacker (defender) can rank the physical components based on the magnitude of their associated $f_p$. Accordingly, each player can maximize (minimize) the harm caused to the system taking into consideration the defense (attack) strategy that can be adopted by the opponent. However, cyber-physical systems are known to be very complex systems. Thus, obtaining such an exact ranking of the costs caused by the loss of every component, $f_p$, is highly complex. As a result, in practice, the attacker and defender can build their own perception of the vector of incurred losses $\boldsymbol{f}$, denoted as $\hat{\boldsymbol{f}}^i$ for $i\in\mathcal{I}$, then take action accordingly. However, $\hat{\boldsymbol{f}}^i$ can differ from $\boldsymbol{f}$. Moreover, the attacker and defender can have different computational capabilities and thus can generate a different $\hat{\boldsymbol{f}}^i$ based on their skill and computational levels. Thus, by following its own perception, a player can deviate from full rationality while choosing its optimal strategy. Consequently, %In this case, a player is known to act with 
this bounded rationality %~\cite{Boundedrationality} which 
can lead to deviations from the NE strategies. 

To model such bounded rationality, we categorize each player based on its \emph{level of thinking} which is defined by how close is its perception $\hat{\boldsymbol{f}}^i$ to the actual $\boldsymbol{f}$. Thus, high level thinkers are more intelligent, have better knowledge, and superior computational ability, allowing them to generate a closer perception to the real $\boldsymbol{f}$. Such a notion is inspired from the behavioral framework of \emph{cognitive hierarchy} (CH)~\cite{CH01} in which it is shown that human players assume that they have the highest level of thinking, denoted by level $K$, and that their opponents' levels of thinking are distributed over lower levels $0,...,K-1$. In a CH model, level $0$ thinkers choose an action randomly from their strategy space while higher level thinkers employ more advanced levels of reasoning to choose their strategies. To model the proportion of level $k$ thinkers for $k\in\{0,...,K-1\}$, a Poisson distribution, $\alpha(k)$, with mean and variance denoted by $\lambda$ is usually assumed~\cite{CH01}:
\begin{align}\label{eq:Poisson}
\alpha(k)=\frac{e^{-\lambda}\lambda^k}{k!}.
\end{align}

Given that the defender in our model is the system operator, it can anticipate with full certainty $\boldsymbol{f}$. In contrast, the attacker might have a distorted $\hat{\boldsymbol{f}}$ which reflects its level of thinking and, as a result, its chosen strategy. The defender, hence, has to choose its optimal strategy while anticipating the possibility of facing an attacker that can fall in any category $k$ with a probability $\alpha(k)$.   

As a result, the knowledge that the defender has about potential types of the attacker that it can face can give an incentive for the defender to deviate from its NE strategy. In fact, this anticipation of the various attacker types would change the best response strategy of the defender from the NE strategy which assumes that the defender faces only a fully rational attacker. Moreover, since the attacker acts with bounded rationality based on its thinking level, this attacker chooses the strategy that it perceives to be optimal based on its own perception following from its $\hat{\boldsymbol{f}}$. Thus, the attacker also has an incentive to deviate from the NE strategy. Consequently, our bounded rationality framework showcases how in practical situations attackers and defenders can deviate from the fully rational NE.   

To give more elaboration of our proposed CPS model, game formulation, and bounded rationality framework, we next analyze in detail a case study focusing on the concept of wide area protection of a smart grid with energy markets implications.                                                 
%we use noncooperative game theory~\cite{GT01} to analyze their optimal decision making processes. In particular, we formulate a static, strategic noncooperative game $\Xi=\langle \mathcal{P}, (\mathcal{A}_{i})_{i\in\mathcal{P}}, (U_{i})_{i\in\mathcal{P}}\rangle$, where $\mathcal{P}\triangleq \mathcal{M} \cup \{0\}$ is the players' set, composed of all $M$ attackers and the defender that is referred to via index $0$, $\mathcal{A}_i$ is the set of actions available to player $i \in \mathcal{P}$ which consists of choosing an attack/defense vector, $\boldsymbol{a}_i\in\mathcal{A}_i$, and $U_i$ is the utility function of player $i$. 

%this cyber-physical system is subject to cyber attacks on the cyber nodes which can lead to the loss of physical components leading to an incurred loss to the system. Thus, we consider an attacker which chooses a set of cyber nodes to attack. To this end, when a cyber node $c_k$ is attacked, the probability of failure of that node goes up to 1. In other words, this leads to $\pi_k^c=1$. This will accordingly incur a risk of failure to the physical components that are interconnected to $c_k$.   aiming at incurring the worst loss to the system.  

\section{Wide Area Protection in the Smart Grid}\label{sec:WideAreaProtectionandMarkets}
\subsection{Smart Grid Wide Area Monitoring and Protection}\label{subsec:WideAreaProtection}
\begin{figure}[t!]
  \begin{center}
   \vspace{-0.35cm}
    \includegraphics[width=8cm]{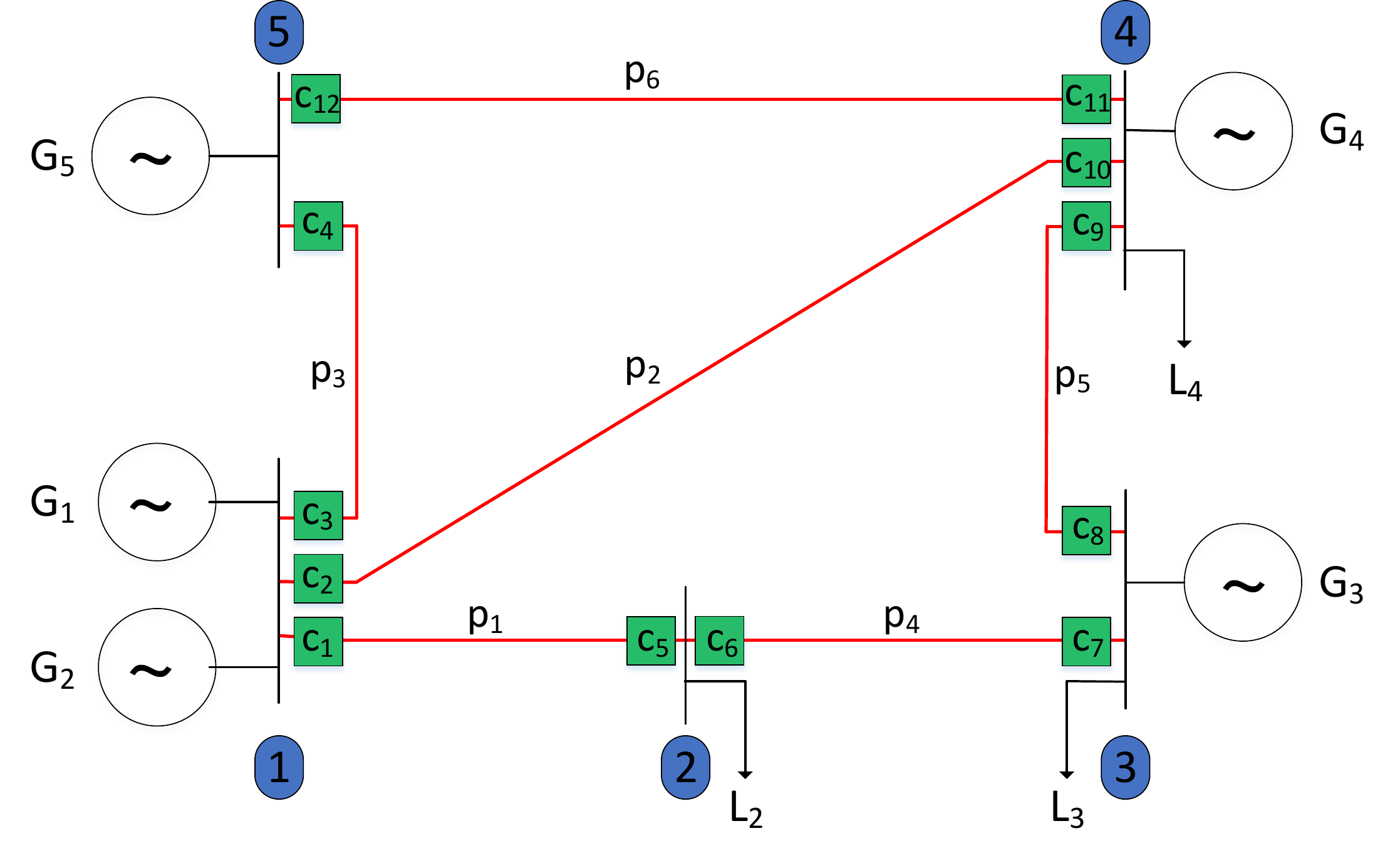}
    \vspace{-0.4cm}
    \caption{\label{fig:PJM5BusSys} PJM 5-bus System}
  \end{center}\vspace{-0.9cm}
\end{figure}
%
%In a smart grid, wide area monitoring, protection, and control apply protective actions to prevent the propagation of large disturbances~\cite{WideAreaPhadke}. This monitoring and protection concept relies on system-wide information sent from a collection of cyber nodes to generate protective actions affecting the status (i.e. connectivity) of the physical components. The extent to which the information sent by every cyber node affects the status of each physical component can follow our proposed model in Section~\ref{sec:CPSSecModel}.
In a smart grid, the concept of wide area monitoring, protection, and control relies on system-wide information sent from a collection of cyber nodes to generate protective actions affecting the status (i.e. connectivity) of the system's physical components to prevent the propagation of large disturbances~\cite{WideAreaPhadke}. The extent to which the information sent by every cyber node affects the status of each physical component can follow our proposed model in Section~\ref{sec:CPSSecModel}.

Consider the PJM 5-bus system shown in Fig.~\ref{fig:PJM5BusSys}. This test system comprises 5 generator units and 3 loads. All data pertaining to this test system are available in~\cite{PJM5Bus}. The cyber nodes $\mathcal{C}\triangleq\{c_1,...,c_{12}\}$ collect real time data from around the system and send them to the SCADA. The SCADA processes the data, detects possible disturbances, and sends, in this event, protection actions requiring the disconnection of a transmission line to stop the propagation of the disturbance. The transmission lines, $\mathcal{P}\triangleq\{p_1,...,p_{6}\}$, constitute the physical nodes of the system.  %a big disturbance to the system that can be caused by a fault. 
%To this end, the SCADA can detect the occurrence of a fault over a transmission line and hence order to disconnect that line from the system. 

Accordingly, one of the purposes of this wide protection concept is to disconnect physical components, such as transmission lines, to stop the propagation of a detected disturbance. This is known as disturbance isolation. When the protection system successfully isolates a disturbance, as per its design requirements, the underlying protection scheme is known to be ``dependable''~\cite{PhadkeRelayingBook}. In addition, the protection system is required to be ``secure'' which dictates that the system takes protective actions \emph{only} in the event of occurrence of anomalies~\cite{PhadkeRelayingBook}. Thus, falsely disconnecting a component of the system during normal operation is seen as a security breach. %to the security of that system. %Power system protection engineers are typically faced with the challenge of biasing the protection system towards making it more dependable or more secure since these two design criteria are often contradictory.  

To this end, a malicious attacker can target the security of the system by compromising a number of cyber nodes $n_a$ and manipulating their sent data, to falsely trip a certain transmission line. %Following from our description, an attacked measurement is considered to have failed with a probability equals to 1.  in 
Here, $\boldsymbol{\kappa}=[\kappa_1,...,\kappa_{12}]$ is the failure probability vector of the 12 cyber nodes (i.e. probability of a cyber node sending false data) and $\boldsymbol{\pi}=[\pi_1,...,\pi_6]$ is the vector of probabilities of a false disconnection of a transmission line due to one, or multiple, failures in the cyber system. 
The degree up to which a failure on the cyber side leads to a disconnection of a transmission line is captured by the matrix $\boldsymbol{R}$ in~(\ref{eq:RiskMatrix}). Locally collected data give, naturally, a better indication of the real-time operating state of a transmission line and hence have the most significant effect on the decision of disconnecting that line. Based on this observation, we build $\boldsymbol{R}$ as follows. 
As shown in Fig.~\ref{fig:PJM5BusSys}, each transmission line is affected by data sent from 12 cyber nodes 2 of which are locally connected to it. These local cyber nodes equally share a 50\% effect on the decision to disconnect the line while the other 50\% is split equally between the 10 remaining cyber nodes.  As a result, $\boldsymbol{R}$, such that $\boldsymbol{\pi}=\boldsymbol{\kappa}\boldsymbol{R}$, 
is represented as follows\footnote{We use this representation of $\boldsymbol{R}$ as a numerical example of our proposed model. Nonetheless, other numerical representations could have also been equally adopted without affecting the validity of our model and underlying analyses.}: 
\begin{numcases}
{r_{i,j}=}
0.25, \>\>\textrm{if}\> c_i \>\textrm{is locally connected to}\> p_j, \nonumber \\ \nonumber
0.05, \>\>\textrm{otherwise}.
\end{numcases}
%\begin{align}
%\small
%\boldsymbol{R}=
%\left( \begin{array}{cccccc}
%0.25 & 0.05 & 0.05 & 0.05 & 0.05 & 0.05 \\ %1
%0.05 & 0.25 & 0.05 & 0.05 & 0.05 & 0.05 \\ %2
%0.05 & 0.05 & 0.25 & 0.05 & 0.05 & 0.05 \\ %3
%0.05 & 0.05 & 0.25 & 0.05 & 0.05 & 0.05 \\ %4
%0.25 & 0.05 & 0.05 & 0.05 & 0.05 & 0.05 \\ %5
%0.05 & 0.05 & 0.05 & 0.25 & 0.05 & 0.05 \\ %6
%0.05 & 0.05 & 0.05 & 0.25 & 0.05 & 0.05 \\ %7
%0.05 & 0.05 & 0.05 & 0.05 & 0.25 & 0.05 \\ %8
%0.05 & 0.05 & 0.05 & 0.05 & 0.25 & 0.05 \\ %9
%0.05 & 0.25 & 0.05 & 0.05 & 0.05 & 0.05 \\ %10
%0.05 & 0.05 & 0.05 & 0.05 & 0.05 & 0.25 \\ %11
%0.05 & 0.05 & 0.05 & 0.05 & 0.05 & 0.25 \\ %12
%\end{array} \right)   
%\normalsize
%\end{align}  

Next, we focus on the cost of loss, $f_{p_i}$, of each physical component $p_i\in\mathcal{P}$. A wrongful disconnection of a transmission line can have detrimental effects on the operation and stability of a power system. For example, the 1965 blackout of the Northeast region of the United States and the Ontario province of Canada was caused by a false trip of a transmission line% by a misconfiguration of a protective relay on one of the transmission lines. 
%This relay, which % is designed to trip for a current that exceeds a given threshold was set too low. This 
%was set too low, %has led to a false disconnection of that 
%falsely disconnected a tranmission line which resulted in a cascading chain of events that ended up in a wide scale blackout. This blackout left around 30 million people in Ontario, Canada and 8 U.S states with no power for up to 13 hours
~\cite{TheGridSchewe2007}. As a result of such incidents, power system operators adopted what is known as the ``$n-1$ security criterion'' which requires the system to preserve its normal state of operation after the loss of one of its $n$ components~\cite{woodwollenberg}. Based on this reinforced security requirement, a loss of one transmission line does not, typically, affect the safety of a system under low stress  operating conditions. Thus, we will focus on another key effect of a false disconnection of a transmission line, namely, the economic effect.% \footnote{The power system's response to changes in system conditions follows 3 time scales: 1) governor response (first few seconds), 2) power frequency control (few more seconds), and lastly, 3) economic dispatch (next few minutes)~\cite{HierarchicalPS}. We focus our case analysis on the longer time scale. Nonetheless, analysis of shorter time scale controls is an interesting extension of our work that would be targeted in future research.}  

%\subsection{Energy Markets}\label{subsec:EnergyMarkets}    
%
%As depicted in~\cite{woodwollenberg,HierarchicalPS}, the controlled response of a power system to changing system conditions follows a time-scale hierarchy. This hierarchy is based on a time decomposition that results from the different implemented control levels' response times. In fact, after a protection action is performed, the system's response follows three time scales. On the first time scale (first few seconds), governor control at every power plant takes action to compensate for the change in frequency due to loading changes in the system. On the second time scale (for the following few seconds), load-frequency control schemes control the power flow through the tie lines of the various interconnected systems. On the third time level (following few minutes), economic dispatch of the system is performed to run the system in the most economical manner after disturbance occurrence and elimination. We turn our focus in our case analysis on the longer time scale to analyze the effects on the economic dispatch of a smart grid caused by malicious attacks on cyber nodes leading to the disconnection of transmission line(s)\footnote{The effect of this disconnection on shorter time scale controls is an interesting extension of our work that would be targeted in future research works.}. 
The economic dispatch of the smart grid is based on the solution of an optimal power flow (OPF) problem. A typical OPF problem formulation~\cite{OPF} is an optimization problem aiming at minimizing the total generation cost of the system subject to a set of equality and inequality constraints reflecting the system's operational requirements and physical limits. %the power flow equations and inequality constraints reflecting bus voltage limits, line flow limits: 1) the power flow balance equations (real and reactive power entering each bus are equal to real and reactive power leaving it), 2) power output limits for every generator, 3) voltage limits on every bus, 4) power flow limits over every transmission line, etc. As a result, the loss of a transmission line affects the obtained optimal solution.

To this end, consider $V^0$ to be the value function of the original OPF problem, i.e. without loss of any transmission line, and $V^{p_i}$ to be the value function of the OPF with loss of transmission line $p_i\in\mathcal{P}$. The value function of the OPF problem reflects the total cost of generation spent to meet the load and is normally expressed in $\$$ per hour. Moreover, consider $T^{p_i}$ (expressed in hours) to be the time needed to bring back $p_i$ into operation and $CR^{p_i}$ (expressed in $\$$) to be the cost of repair of $p_i$. Then, $f_{p_i}$ will be given by:
\begin{align}\label{eq:fip}
f_{p_i}=(V^{p_i}-V^0)T^{p_i}+CR^{p_i}.
\end{align} \vspace{-0.6cm} 
  
\subsection{Numerical Results}\label{subsec:NumericalResults}  
\begin{figure}[t!]
  \begin{center}
   \vspace{-0.35cm}
    \includegraphics[width=8cm]{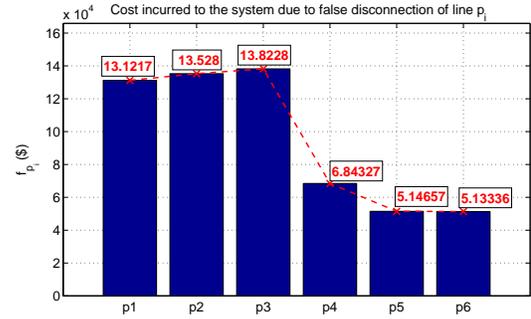}
    \vspace{-0.6cm}
    \caption{\label{fig:LossOfLineCost} Cost to the system incurred by the loss of each transmission line as expressed in~(\ref{eq:fip}).}
  \end{center}\vspace{-0.9cm}
\end{figure}

For the considered PJM 5-bus system, to calculate $f_{p_i} \> \forall p_i\in\mathcal{P}$, we run the optimal power flow seven different times to compute $\{V_0, V^{p_1},...,V^{p_6}\}$. Also, we consider that every disconnected transmission line needs 12 hours and costs $\$80,000$ to be brought back into operation\footnote{Such numbers are used as an example and are usually specific to the system and to the line's voltage level and length.}. %~\cite{CostOfTransmissionLine}.}. %, i.e. $T^{p_i}=12 h$ and $CR^{p_i}=\$80,000 \, \forall p_i=\{p_1,...,p_6\}$. 
The results are shown in Fig.~\ref{fig:LossOfLineCost}.% where the $f^p_i$ values in red have units of $\$10,000$. % as indicated by the vertical axis scale. 
As can be seen from Fig.~\ref{fig:LossOfLineCost}, disconnecting $p_3$ incurs the highest cost to the system ($\$131,220$) followed by $p_2, p_1, p_4, p_5$ and $p_6$, respectively. 

%As discussed in Section~\ref{subsec:GameFormulation}, we consider a game between an attacker and a defender. In this game, the attacker/defender chooses a number of cyber nodes to attack/defend. % while the defender chooses a number of cyber nodes to defend. 
%When $c_j$ is attacked, $\pi_j^c=1$ while when $c_j$ is defended, $\pi_j^c=0$ even when $c_j$ is attacked. 
In our case analysis, we consider that the attacker (defender) aims at attacking (defending) a given transmission line, $p_i\in\mathcal{P}$, by compromising (securing) the two cyber nodes that have the most effect on this line. In other words, for the considered game, the strategy space of the defender and attacker can be defined as follows: $\mathcal{S}_d=\,\mathcal{S}_a=\{(c_1,c_5), (c_2,c_{10}), (c_3,c_4), (c_6,c_7), (c_8,c_9), (c_{11},c_{12})\}$.
%\begin{flalign}\label{eq:AttDefSpaces}
%\mathcal{S}^d&=\,\mathcal{S}^a=\\ \nonumber
%&\{(c_1,c_5), (c_2,c_{10}), (c_3,c_4), (c_6,c_7), (c_8,c_9), (c_{11},c_{12})\},
%\end{flalign}
This corresponds to choosing to defend/attack one of the lines in $\mathcal{P}$. Thus, equivalently, $\mathcal{S}_d=\mathcal{S}_a=\{p_1,p_2,p_3,p_4,p_5,p_6\}$. 

The defender and attacker aim at maximizing their expected utility functions, $\bar{U}_i$, given by~(\ref{eq:ExpectedU}). 
%\begin{flalign}\label{eq:EUT}
%\mathbb{E}_{d}(\boldsymbol{\gamma}^d,\boldsymbol{\gamma}^a)&=-\mathbb{E}_{a}(\boldsymbol{\gamma}^d,\boldsymbol{\gamma}^a)\nonumber\\
%&=-\sum_{s^d_i=1}^{|\mathcal{S}_d|}\sum_{s^a_j=1}^{|\mathcal{S}_a|}\gamma^d(s^d_i)\gamma^a(s^a_j)U_a(s^d_i,s^a_j),  
%\end{flalign}
The payoff $U_d(s_d,s_a)$ of the defender for the different $s_d\in\mathcal{S}_d$ and $s_a\in\mathcal{S}_a$ is presented in Table~\ref{table:AttackerPayoff} in which the row player is the defender and the column player is the attacker\footnote{The defender always has a negative payoff even when the attack is blocked. In fact, with no attack, $\boldsymbol{\pi}^c\ne\boldsymbol{0}$ since cyber node $c_k$ can fail due to a software bug or misconfiguration. In our numerical analysis we take $\pi^c_j=1/12 \,\forall j=1,...,12$ prior to attack and defense. $\boldsymbol{\pi}^c$ is small but not equal $\boldsymbol{0}$.}. 
\begin{table}[t!]
\caption{Attacker's Payoff $U_a(s^d_i,s^a_j)$, [Unit: $\$1,000$] \label{table:AttackerPayoff}} \vspace{-0.55cm}
%
%%\vspace{-0.4cm}
\begin{center}
%%\footnotesize
\begin{adjustbox}{max width=\textwidth}
\begin{tabular}[b]{|l|c|c|c|c|c|c|}
\hline
%$p_1$ & $p_2$ & $p_3$ & $p_4$& $p_5$ & $p_6$ \\
%\hline
%\multicolumn{3}{|c|}{For defense strategy $a_\textrm{0}\triangleq a_{0,\textrm{no}}$} \\
%\hline
%
\backslashbox{$d$}{$a$} & $p_1$ & $p_2$ & $p_3$ & $p_4$& $p_5$ & $p_6$ \\
\hline
$p_1$ & -38.82	& -141.22	&-142.30	&-116.71	&-110.49 &	-110.44 \\
\hline
$p_2$ &-139.60	&-38.69&	-142.17	&-116.58	&-110.35	&-110.31 \\
\hline
$p_3$ &-139.50	&-140.99&	-38.59&	-116.48&	-110.26&	-110.21 \\
\hline
$p_4$ &-141.82	&-143.31&	-144.40&	-40.92&	-112.58&	-112.53 \\
\hline
$p_5$ &-142.39	&-143.88&	-144.96&	-119.37&	-41.48&	-113.10 \\
\hline
$p_6$ &-142.39&	-143.88&	-144.97&	-119.37	&-113.15& -41.49 \\
\hline
\end{tabular}
\end{adjustbox}
\end{center}
\vspace{-0.85cm}
%%\vspace{-0.5cm}
\end{table}
%%%%%%%%%%%%%%%%%%%%%%%%%%%%%%%
%%% Include Figure
%%%%%%%%%%%%%%%%%%%%%%%%%%%%%%%
%\begin{figure}[t!]
%  \begin{center}
%   \vspace{-0.35cm}
%    \includegraphics[width=7.8cm]{LMPsCase1Paper.eps}
%    \vspace{-0.55cm}
%    \caption{\label{fig:LMPscase1Paper} System LMPs under different attack scenarios.}
%  \end{center}\vspace{-0.9cm}
%\end{figure}
%%%%%%%%%%%%%%%%%%%%%%%%%%%%%%%
The payoff of the attacker for the different strategy combinations is the negative of that of the defender given that the game is of zero-sum type. 

When both players are strategic thinkers and have complete information of the game, the NE in~(\ref{eq:MSNE}) can be found using the von Neumann indifference principle~\cite{GT01}. %as follows:. %This indifference principle is based on the fact that for a player to randomize over its strategy set, the payoffs it obtains from playing each of its strategies, facing a mixed strategy of the opponent, need to be identical. Otherwise, this player would not randomize and it would rather choose the strategy that generates its largest payoff, faced with any mixed strategy of the opponent, with probability $1$. 
Applying the von Neumann indifference principle we get the following equilibrium results:
\begin{align}
\boldsymbol{\gamma}^*_{d}=[0.2931, 0.3034, 0.3107,  0.0842,  0.0047,  0.0040], \label{eq:AttNE}\\
\boldsymbol{\gamma}^*_{a}=[0.1276, 0.1244, 0.1222, 0.1922,  0.2167,  0.2169].
\end{align}  
This optimal strategy leads to $\bar{U}_{d}=-\$110,240$ and $\bar{U}_{a}=\$110,240$.

On the other hand, as explained previously, the NE assumes that both attacker and defender act strategically and have complete information of the game. However, computing $\boldsymbol{f}$ requires the solution of an OPF which is a complicated optimization problem~\cite{StottOPF} that requires complete knowledge of the system. The complexity of finding the solution of the OPF in practical applications is thoroughly discussed in~\cite{StottOPF}. To this end, since the defender is the system operator, it has complete knowledge of the system and has the computational tools that are developed specifically for the solution of the system's OPF. On the other hand, the attacker might not have neither the full knowledge of the system nor the computational capabilities to solve the OPF and compute $f_{p_i} \, \forall p_i\in\mathcal{P}$. In this case, the attacker must build a perception of the ranking of $f_{p_i}$, for different $p_i\in\mathcal{P}$, to assess which attack strategy is the most harmful to the system. Thus, playing an NE defense strategy against an assumed fully rational attacker might not be an optimal strategy given that the attacker can deviate from its NE strategy due to its bounded rationality. %As a result, the attacker would base its attack on its perceived ranking. 
%Since this perceived ranking might not coincide with the true $f^p_i$ ranking, the rationality of the attacker when choosing its attack strategy is considered to be limited. 

By applying the proposed model of Section~\ref{subsec:BoundedRationality}, we can investigate deviations from the NE due to the bounded rationality of the attacker. To this end, we consider that the attacker can take one of three types reflecting three different ``thinking levels'' as described in Section~\ref{subsec:BoundedRationality}. 
A \emph{level 0} attacker, denoted by $l_0$, is one that chooses an attack strategy randomly (following a uniform distribution) from its strategy set $\mathcal{S}_a$. 
A \emph{level 1} attacker, denoted by $l_1$, cannot generate OPF solutions but can observe the power flow on each line (requires eavesdropping rather than solving the computationally demanding OPF). Hence, an attacker $l_1$ builds a perception of the most harmful line to attack based on the level of power flow on every line. A more loaded line, $p_i$, is associated with the largest $\hat{f}_{p_i}$. Thus, an $l_1$ attacker targets the line that is the most loaded. %Here we note that $l^1$ performs a higher thinking level than $k^0$ since it is considered to have access to the line flow data in the system. 
A \emph{level 2} attacker, denoted by $l_2$, is considered to have full knowledge of the system and high computational ability and can hence solve the OPF and compute $f_{p_i} \, \forall p_i\in\mathcal{P}$. % which it can acquire through hacking the operator's computer system and extracting such information. %However, $k_2$ is still not able to re-solve the OPF problem to obtain $f^p_i$ associated with the loss of every $p_i$. 
Thus, $l_2$ can compute the exact $\boldsymbol{f}$ and attacks the line $p_i$ with highest $f_{p_i}$. 

In our model, the defender performs the highest thinking level since it has the capability and knowledge to think strategically. In fact, through historical data, the defender can build an anticipation about the potential thinking levels that an attacker may perform. Thus, the defender anticipates what the attack strategy can be, based on a distribution of possible attacker's types, and plays a best response defense strategy that maximizes its expected payoff. 
On the other hand, the attacker may not be able to acquire such accurate knowledge about what the defender's strategy may be. Thus, the attacker forms a perception of the harm that its attack can have. Then, the attacker bases its attack on this perception since it assumes that the defender is equally likely to defend any of the cyber nodes. In other words, the attacker assumes the defender to be a level 0 thinker. %On the other hand, the defender is a strategic thinker that considers facing an attacker that can be of any of the 3 types and selects an optimal strategy that returns the highest possible expected payoff considering potential actions of the attacker. 

Next, we compute the best response strategy of the defender when faced with an attacker belonging to one of the three types. $\gamma_a^{l_k}$ corresponds to attacker $l_k$'s attack strategy while $\gamma_d^{l_k}$ denotes the best response of the defender against this strategy.  

\begin{figure}[t!]
  \begin{center}
   \vspace{-0.35cm}
    \includegraphics[width=8cm]{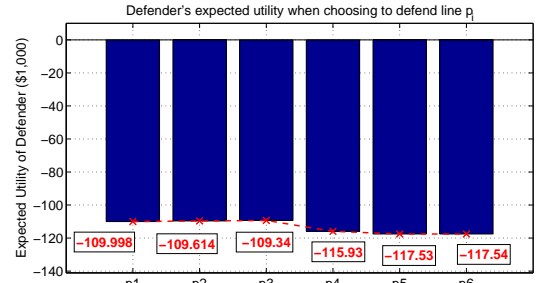}
    \vspace{-0.6cm}
    \caption{\label{fig:DefenderVsl0Attacker} Defender's expected utility when defending one of the lines $p_i\in\mathcal{P}$ facing an $l_0$ attacker.}
  \end{center}\vspace{-0.9cm}
\end{figure}

To this end, we first consider $l_0$ which chooses a line to attack randomly. Thus, its strategy is given by $\gamma_a^{l_0}(s_a)=1/6 \,\forall s_a\in\mathcal{S}_a$. %The best response strategy to the defender facing strategy $\gamma^a_{l^k}$ of an attacker of type $l^k$ is denoted by $\gamma^d_{l^k}$ and is one that solves:
%\begin{align}
%\max_{\gamma^d} \mathbb{E}_{U_d}(\gamma^d,\gamma^a_{l^k}).
%\end{align}
To determine the best response of the defender facing an $l_0$ attacker, we show, in Fig.~\ref{fig:DefenderVsl0Attacker}, the expected utility of the defender when choosing each of its possible strategies. By checking the values of the achieved expected utility when defending a line $p_i$ (dashed line in Fig.~\ref{fig:DefenderVsl0Attacker}), one can see that the best response of the defender against an $l_0$ attacker is to choose to defend $p_3$. That is, $\gamma_d^{l_0}(p_3)=1$ and $\gamma_d^{l_0}(p_j)=0 \,\forall p_j\ne p_3$  
which results in $\bar{U}_{d}^{l_0}(\gamma_d^{l_0},\gamma_a^{l_0})=-\bar{U}_{a}^{l_0}(\gamma_d^{l_0},\gamma_a^{l_0})=-\$109,340$.  

Considering the case of an $l_1$ attacker, its bounded rationality dictates to attack the line carrying the highest power flow since a disconnection of such a line is perceived to cause the highest harm to the system. 
Let $\boldsymbol{\omega}$ be the vector of power flows over lines $\{p_1,...,p_6\}$ with no disconnection of any of these lines. Running an OPF of the PJM 5-bus system results in the following flows expressed in $MW$: $\boldsymbol{\omega}=[252.38,  187.87, 230.25,  49.21,  24.95, 238.5]$. Given that $p_1$ bears the highest power flow, $\gamma_a^{l_1}$ consists of attacking line $p_1$ with probability equals to 1. Based on Table~\ref{table:AttackerPayoff}, the defender's best response against $\gamma_a^{l_1}$ is to choose to defend line $p_1$ with probability 1 ($\gamma_d^{l_1}(p_1)=1$). These defense and attack strategies result in $\bar{U}_{d}^{l_1}(\gamma_d^{l_1},\gamma_a^{l_1})=-\bar{U}_{a}^{l_1}(\gamma_d^{l_1},\gamma_a^{l_1})=-\$38,830$.  

In contrast to $l_1$, an $l_2$ attacker has the ability and the knowledge to solve the OPF and characterize the line with highest $f_{p_i}$. As shown in Fig.~\ref{fig:LossOfLineCost}, line $p_3$'s loss is the most harmful. Hence, $\gamma_a^{l_2}$ consists of attacking line $p_3$ with probability 1. The best response of the defender to this strategy can be obtained from Table~\ref{table:AttackerPayoff} and consists of defending $p_3$ with a probability 1 ($\gamma_d^{l_2}(p_3)=1$). These defense and attack strategies result in $\bar{U}_{d}^{l_2}(\gamma_d^{l_2},\gamma_a^{l_2})=-\bar{U}_{a}^{l_2}(\gamma_d^{l_2},\gamma_a^{l_2})=-\$38,590$.  

Given that the defender might be faced with an attacker from any of the three types, it aims to devise an optimal strategy that achieves the best expected utility facing the possible three types. The probability that the attacker is of level $l_k$ is given by $\alpha(k)$ in~(\ref{eq:Poisson}). As seen from~(\ref{eq:Poisson}), the ratio of probabilities of level $k+1$ to level $k$ is a constant that we denote by $\tau$. Thus, $\alpha(1)/\alpha(0)=\alpha(2)/\alpha(1)=\tau$. Using this relation and noting that $\alpha(0)+\alpha(1)+\alpha(2)=1$, we can express $\alpha(0)$ as: %in terms of $\tau$:
%\begin{align}\label{eq:alpha0tau}
$\alpha(0)=1/(1+\tau+\tau^2)$.
%\end{align} 
 
From our derived best response expressions, $\gamma_d^{l_0},\, \gamma_d^{l_1}$ and $\gamma_d^{l_2}$, we know that the defender would defend line $p_3$ when faced with an $l_0$ or $l_2$ attacker while the defender would defend line $p_1$ when faced with an $l_1$ attacker. 
Given the defined probabilities of each attacker's type, we can calculate the expected payoff of the defender when faced with an attacker $l_k\triangleq\alpha(0)l_0+\alpha(1)l_1+\alpha(2)l_2$. This latter notation means that $l_k$ corresponds to a combination of types $l_0, l_1$ and $l_2$ with probability $\alpha(0), \alpha(1)$ and $\alpha(2)$, respectively.  Following from the expressions of $\bar{U}_{d}^{l_0}, \,\bar{U}_{d}^{l_1}$ and $\bar{U}_{d}^{l_2}$ as well as from Table~\ref{table:AttackerPayoff}, the expected utility of the defender when defending $p_1$ and $p_3$ can be expressed as follows: \vspace{-0.25cm}
\begin{flalign}
\bar{U}_{d}(p_1,l_k)&=\alpha(0)\bar{U}_{d}(p_1,l_0)+\alpha(1)\bar{U}_{d}(p_1,l_1)+\alpha(2)\bar{U}_{d}(p_1,l_2)\nonumber\\
&=-109.998\alpha(0)-38.823\alpha(1)-142.302\alpha(2) \nonumber \\
&=-109.998\alpha(0)-38.823\alpha(0)\tau-142.302\alpha(0)\tau^2 \nonumber\\
\bar{U}_{d}(p_3,l_k)&=-109.336\alpha(0)-139.489\alpha(1)-38.823\alpha(2) \nonumber \\
&=-109.336\alpha(0)-139.489\alpha(0)\tau-38.823\alpha(0)\tau^2 \nonumber  
\end{flalign} 
\vspace{-0.6cm}%\\

As a result, the defender picks $p_1$ when $\bar{U}_{d}(p_1,l_k)>\bar{U}_{d}(p_3,l_k)$, picks $p_3$ when $\bar{U}_{d}(p_1,l_k)<\bar{U}_{d}(p_3,l_k)$, and is indifferent when $\bar{U}_{d}(p_1,l_k)=\bar{U}_{d}(p_3,l_k)$. Thus, the defender chooses the strategy, $\gamma_{d}^{l_k*}$ which results in $\bar{U}_d^*(\gamma_{d}^{l_k*},l_k)=\textrm{max}\big(\bar{U}_{d}(p_1,l_k),\bar{U}_{d}(p_3,l_k)\big)$. 
\begin{figure}[t!]
  \begin{center}
   \vspace{-0.35cm}
    \includegraphics[width=8cm]{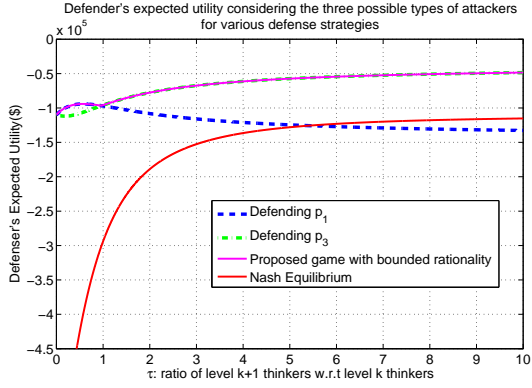}
    \vspace{-0.4cm}
    \caption{\label{fig:DefenderEUTNEvsCH} Defender's expected utility when facing an attacker which can be of types $l_0$, $l_1$, and $l_2$.}
  \end{center}\vspace{-0.9cm}
\end{figure} 

Fig.~\ref{fig:DefenderEUTNEvsCH} shows the optimal expected utility achieved by the defender when playing $\gamma_{d}^{l_k*}$ for an increasing ratio $\tau$. %it considers facing an attacker with multiple possible types for an increasing ratio $\tau$.  
Fig.~\ref{fig:DefenderEUTNEvsCH} shows that the defender achieves a better expected utility, $\bar{U}_{d}^*$, when playing $\gamma_{d}^{l_k*}$ against $\gamma_a^{l_k}$ as compared to the NE utility achieved when choosing $\gamma_d^*$, in~(\ref{eq:AttNE}), against $\gamma_a^{l_k}$. Thus, given that the attacker can act with bounded rationality in security applications, accounting for this bounded rationality has achieved a better payoff for the defender as compared to playing the NE strategy. For $\tau=0.5$, the defender achieves a $78\%$ increase in its expected utility by choosing  $\gamma_{d}^{l_k*}$ instead of $\gamma_{d}^*$. This increase drops to $67\%$ for $\tau=1$ and $55\%$ for $\tau=5$. The value of $\tau$ gives an indication about the probability of having a lower or higher level attacker. In fact, $\tau<1$ indicates that a low level attacker is more probable while $\tau>1$ indicates that a higher level attacker is more probable. Thus, the general trend shows that when the probability of a high level attacker increases, the gain from deviating from the NE defense strategy decreases. %In fact, for $\tau=0.5$, i.e. a lower level attacker is more probable, the defender achieves an increase of $\%14.5$ under the bounded rationality framework, $\mathbb{E}_{d}^{BR}$, as compared to the NE solution, $\mathbb{E}_{d}^{NE}$. For $\tau=1$, i.e. attacker has equal probability of being any of the 3 types, the defender's $\mathbb{E}_{d}^{BR}$ achieves an increase of $13\%$ over the $\mathbb{E}_d^{NE}$. Equally, for greater values of $\tau$, namely, $\tau=2$ and $\tau=5$, i.e. a higher level attacker is more probable, the defender's $\mathbb{E}_{d}^{BR}$ achieves an increase of respectively $30\%$ and $50\%$ over the $\mathbb{E}_d^{NE}$. 

Moreover, it can be seen from Fig.~\ref{fig:DefenderEUTNEvsCH}, that the defender's optimal strategy is to defend $p_1$ for approximately $\tau<1$ and defend $p_3$ for $\tau>1$. This implies that when it is more probable to face a low level attacker, the defender optimally defends against the targeted line, $p_1$. In contrast, when facing a more intelligent attacker is more probable, $\tau>1$, the defender's optimal strategy is to defend $p_3$ which is the most probable target of a high level attacker.       

\section{Conclusion}\label{sec:Conclusion}
%\vspace{-0.08cm}
In this paper, we have studied the security of the smart grid in the presence of an attacker and defender. We have first introduced a general CPS security model showing how attacks can propagate from the cyber to the physical system. We have then formulated the interaction between attacker and defender using a game-theoretic model. In addition, we have introduced a bounded rationality framework inspired by cognitive hierarchy theory that is suitable to model the limited levels of thinking of the attacker. We have applied our framework to the concept of wide area protection of the smart grid and its energy markets implications. We have shown that when considering bounded rationality of the attacker, the defender can achieve a better protection of the system. We have also shown that when the thinking level of the attacker increases, the gain from deviating from the NE defense strategy decreases. %We have also shown how the computational capabilities of the attacker can affect its as well as the defender's chosen strategies. %Our work provide a general framework that can be applied to various areas of CPS security including security of smart grids.     
%\vspace{-0.1cm}
\def\baselinestretch{0.9}
\bibliographystyle{IEEEtran}
\bibliography{reference}

\end{document}